\documentclass[10pt, journal, twocolumn]{IEEEtran}
\usepackage{graphicx}
\usepackage{epstopdf}
\usepackage{amssymb}
\usepackage{amsmath}
\usepackage{amsthm}
\usepackage{amsfonts}
\usepackage{textcomp}
\usepackage{gensymb}
\usepackage{url}
\usepackage{cite}
\usepackage{bm}
\usepackage{tabls}
\usepackage{tikz}
\usetikzlibrary{arrows.meta}
\usepackage{steinmetz}
\usepackage{array}
\usepackage{stackengine}
\usepackage[font=normalsize]{subfig}
\usepackage{float}
\usepackage{dsfont}
\usepackage{arydshln}
\usepackage{mathtools}
\usepackage{balance}
\usepackage{varwidth}

\ifCLASSINFOpdf
\else
\fi

\begin{document}
%


%
%
%

\title{Intelligently Wireless Batteryless RF-Powered Reconfigurable Surface}
\author{\small {Iosif Vardakis{$^*$}, Georgios Kotridis{$^*$}, Spyridon Peppas{$^*$},  Konstantinos Skyvalakis, Georgios Vougioukas and Aggelos Bletsas}\\
School of Electrical and Computer Engineering, Technical University of Crete, Chania 73100, Greece\\[-2pt]
{\small {\tt \{ivardakis, gkotridis, speppas, kskyvalakis, gevougioukas\}@isc.tuc.gr}, {\tt aggelos@telecom.tuc.gr}
}
\thanks{\scriptsize$^*$Equal contributors. This work has been submitted to the IEEE for possible publication. Copyright may be transferred without notice, after which this version may no longer be accessible. The research work was supported by the Hellenic Foundation for Research and Innovation (H.F.R.I.) under the ``First Call for H.F.R.I. Research Projects to support Faculty members and Researchers and the Procurement of High-cost research equipment'' (Project Number: 2846).}
}
\maketitle

\begin{abstract}
This work exploits commodity, ultra-low cost, commercial radio frequency identification tags (RFID) as the elements of a reconfigurable surface. Such batteryless tags are powered and controlled by a software-defined (SDR) reader, with properly modified software, so that a source-destination link is assisted, operating at a different carrier frequency.  In terms of theory, the optimal gain and corresponding best element configuration is offered, with tractable polynomial complexity (instead of exponential) in number of elements. In terms of practice, a concrete way to design and prototype a wireless, batteryless, RF-powered, reconfigurable surface is offered and a proof-of-concept is experimentally demonstrated. It is also found that even with perfect channel estimation, the weak nature of backscattered links limits the performance gains, even for large number of surface elements. Impact of channel estimation errors is also studied. Future extensions at various carrier frequencies could be directly accommodated, through simple modifications in the antenna and matching network of each RFID tag/surface element. 
\end{abstract}
\begin{IEEEkeywords}
Backscatter Radio, RFID, Gen2, Reconfigurable Surface.
\end{IEEEkeywords}
\IEEEpeerreviewmaketitle
\section{Introduction}
\label{sec:intro}
Significant interest has been attracted recently on reconfigurable reflecting surfaces, which are viewed as a way to control the environment, with  a large number of passive elements, i.e., without amplification; such surfaces are envisioned  to offer a \emph{focusing} effect (e.g., work in \cite{HuChHuShAlGeZaAlYuChZhRuReDeMe:20}, 
\cite{BjEmOzOzLaEr:20}, 
 and references therein).   
 A small number of experimental testbeds has recently emerged, e.g., work in \cite{ZhuYaxXieLonShaRotGumWenJam:19}, \cite{DaLiWaBiWaMiYaXuTaJiBiShuXuSheYanFanChZhiRenLaj:20},  \cite{ArBa:20}; \cite{ZhuYaxXieLonShaRotGumWenJam:19} offered a $36$-element array, with phase shifters helping endpoints whose line-of-sight (LoS) is blocked. \cite{DaLiWaBiWaMiYaXuTaJiBiShuXuSheYanFanChZhiRenLaj:20} 
developed a reconfigurable intelligent surface (RIS) with $256$ elements exploiting positive intrinsic-negative (PIN) diodes for $2-$bit phase shifting; \cite{ArBa:20} utilized software-controlled, $2$-load RF switches in groups, in order to select different surface configurations with thousands of elements, exploiting feedback from the destination. All offered testbeds so far are based on wired prototypes.

\begin{figure}[!t]
    \centering
	\includegraphics[width=0.9\columnwidth]{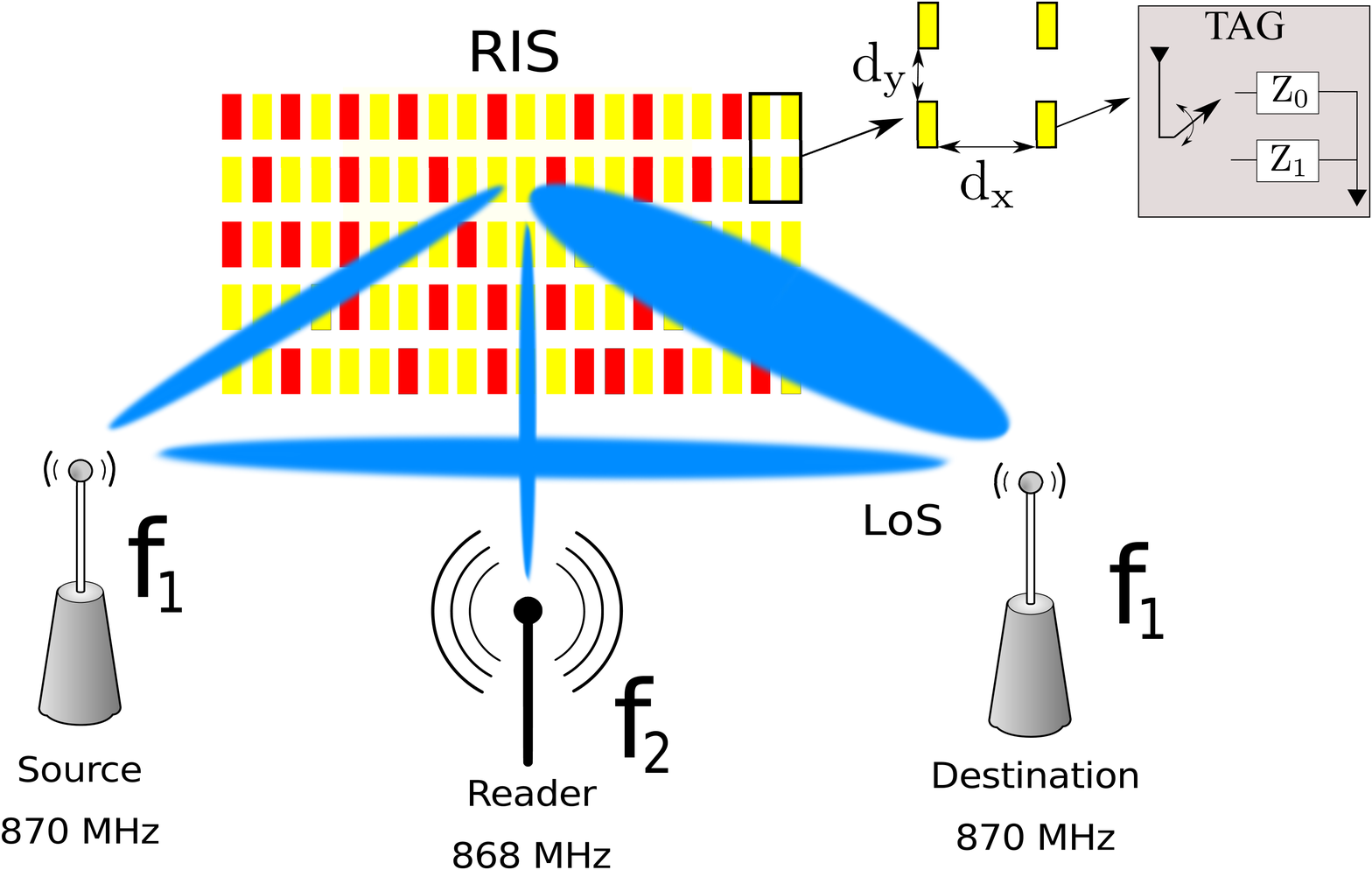}
    \caption{The proposed wireless surface with commodity, wirelessly-controlled, RF-powered, RFID tags.}
    \label{fig:idea}
\end{figure}

Despite the fact that RIS is a special case of bistatic  backscatter radio \cite{KiBlSa:14}, such connection is not widely known in the literature. This work exploits commodity, ultra-low cost, commercial radio frequency identification tags (RFID) \emph{as the RIS elements}. Such batteryless tags are powered and controlled by a software-defined (SDR) reader, operating at carrier frequency $f_2$, with properly modified software, so that a source-destination link, operating at a \emph{different} carrier frequency $f_1$, is assisted (Fig.~\ref{fig:idea}).

It is found for the first time in the literature (to the best of our knowledge): a) the optimal gain and corresponding best RIS element configuration, with tractable polynomial complexity (instead of exponential) in number of elements and b) the way to design and prototype a wireless, batteryless, RF-powered RIS, with commodity RFID tags. Future extensions at various carrier frequencies could be accommodated, through simple modifications in the antenna and matching network of each RFID tag.

\emph{Notation:} $\bm{0}_N$ denotes the all-zeros vector. The phase of complex number $z$ is denoted as $\phase{z}$, while $\Re\{z\}$ denotes the real part 
of $z$. The distribution of a proper complex Gaussian $N \times 1$ vector 
$\bm{x}$ with mean $\bm{\mu}$ and covariance matrix $\bm{\Sigma}$ is denoted by $\mathcal{CN}(\bm{\mu},\bm{\Sigma})\triangleq\frac{1}{\pi^N\det(\Sigma)}\text{e}^{-(\bm{x}-\bm{\mu})^H\bm{\Sigma}^{-1}(\bm{x}-\bm{\mu})}$; the special case of a circularly symmetric complex Gaussian $N \times 1$ vector corresponds by definition to $\mathcal{CN}(\bm{0}_N,\bm{\Sigma})$; 
 expectation of function $\text{g}(\cdot)$ of random variable $x$ is denoted by $\mathbb{E}[\text{g}(x)]$. 
\section{System Model}
\label{sec:System_model}
\subsection{Channel Model}
\label{sec:ch_model}

A source-destination link is assisted by an array of $M$ tags/RIS elements. The following \emph{large-scale} channel path-loss model is adopted \cite{Goldsmith:05}:
\begin{align}
\textsf{L}_\mathrm{X} \propto \left(\frac{\lambda}{4\pi d_0^{\mathrm{X}}}\right)^2\left(\frac{d_0^{\mathrm{X}}}{d_{\mathrm{X}}}\right)^{v_\mathrm{X}},\label{eq:path_loss}
\end{align}
where $\mathrm{X} \in\{\mathrm{SD},\mathrm{ST}_m,\mathrm{T}_m\mathrm{D}\}$ 
denotes the source-to-desti-nation, source-to-tag $m$ and tag $m$-to-destination, respectively; $\lambda$ is the carrier wavelength, $d_0^{\mathrm{X}}$ is a reference distance, $v_{\mathrm{\mathrm{X}}}$ is the path-loss exponent and $d_\mathrm{X}$ is the distance for link $\mathrm{X}$.  

Flat fading is assumed; complex channel coefficient $h_{\text{SD}}$, $h_{\text{ST}_m}$ and $h_{\mathrm{T}_m\mathrm{D}}$ denotes the baseband channel coefficients for the source-destination,  source-tag and tag-reader link, respectively. Due to strong line-of-sight (LoS) signals present in the problem, \emph{small-scale} Rice flat fading channel model \cite{Goldsmith:05} is mainly adopted:\footnote{The complex channel is the superposition of $\sqrt{\frac{\kappa_{\mathrm{T}_m\mathrm{D}}}{\kappa_{\mathrm{T}_m\mathrm{D}}+1}} \,\sigma_{{\mathrm{T}_m\mathrm{D}}} \, e^{j \theta}+\mathcal{CN}\left(0, \frac{\sigma_{{\mathrm{T}_m\mathrm{D}}}^2}{\kappa_{\mathrm{T}_m\mathrm{D}}+1}\right)$ with random $\theta$.}
\begin{align}
h_{\mathrm{T}_m\mathrm{D}}  \sim\mathcal{CN}\left(\sqrt{\frac{\kappa_{\mathrm{T}_m\mathrm{D}}}{\kappa_{\mathrm{T}_m\mathrm{D}}+1}}\,\sigma_{{\mathrm{T}_m\mathrm{D}}},\frac{\sigma_{{\mathrm{T}_m\mathrm{D}}}^2}{\kappa_{\mathrm{T}_m\mathrm{D}}+1}\right),
\end{align}
where $h_{\mathrm{T}_m\mathrm{D}} \stackrel{\triangle}{=} |h_{\mathrm{T}_m\mathrm{D}}|\,e^{-j\phi_{\mathrm{T}_m\mathrm{D}}}$,  $k_{\mathrm{T}_m\mathrm{D}}$ is the power 
ratio between the deterministic LoS component and the scattering components and $\mathbb{E}[|h_{\mathrm{T}_m\mathrm{D}}|^2]=\sigma_{h_{\mathrm{T}_m\mathrm{D}}}^2$ is the average power of the scattering components. For link budget normalization purposes, $\sigma_{h_{\mathrm{T}_m\mathrm{D}}}^2=1$ will be also assumed (other values could be easily accommodated into the large-scale, average coefficients). Similar notation and assumptions hold for $h_{\text{ST}_m}$, $m \in \{1,2, \ldots, M\}$ and $h_{\text{SD}}$. It is noted that for $\kappa = 0$, Rice is simplified to Rayleigh fading. 

Quasi-static block fading is assumed, i.e., the channel remains constant for $L_c$ (source-destination link) symbols and changes independently between channel coherence time periods. Channel coefficients $h_{\text{SD}}$, $\{h_{\text{ST}_m}\}$, $\{h_{\mathrm{T}_m\mathrm{D}}\}$, $m \in \{1,2, \ldots, M\}$ are assumed independent in the numerical results. Furthermore, the following notation is also adopted:
\begin{align}
h_m &=h_{\text{ST}_m}\,h_{\mathrm{T}_m\mathrm{D}}=|h_{\text{ST}_m}\,h_{\mathrm{T}_m\mathrm{D}}|\, e^{-j\phi_m}, m \in \{1,2, \ldots, M\}, \nonumber \\
h_0  &= h_{\text{SD}}, m=0, 
\label{eq:hm}
\end{align}
%

%
%
%

\subsection{Signal Model}
\label{sec:signal_model}
The baseband source message $m(t)$  is given by:
\begin{align}
c(t)= \sqrt{2P} \, m(t)
\end{align}
where $\mathbb{E}[|m(t)|^2]=1$. Different normalization could be incorporated into the large-scale coefficients. The baseband complex equivalent of the scattered waveform from tag $m$ is given by \cite{KiBlSa:14}: 
\begin{align}
u_m(t)&=\sqrt{\eta\,\textsf{L}_{\text{ST}_m}} \, \left[A_s-\Gamma_m(t)\right] \, h_{\text{ST}_m} \, c(t),\\
\Gamma_m(t) & \in \{ \Gamma_1, \Gamma_2, \ldots ,\Gamma_K \},
\end{align}
where $\Gamma_m(t)$ stands for the modified (complex) reflection coefficient for tag $m$, assuming that the tag can terminate its antenna between $K$ loads and $\eta$ models the (limited) tag power scattering efficiency. It is noted that for passive (amplification-free) tags, $|\Gamma_k | \leq 1$, while for commercial RFID tags, $K=2$. Parameter $A_s$ stands for the load-independent \emph{structural mode} that solely depends on tag's antenna \cite{BleDiSah:10}, commonly overlooked in the literature; $A_s = 0$ only for \emph{minimum scattering} antennas, i.e., antennas that do not reflect anything when terminated at open (i.e., infinite) load.

The received demodulated complex baseband signal at the destination is given by the superposition of the source and all tags' backscattered signals propagated 
through wireless channels $h_{\mathrm{SD}}$ and $\{h_{\mathrm{T}_m\mathrm{D}}\}$, respectively:
\begin{align}
&y(t)=\sqrt{\textsf{L}_{\mathrm{SD}}}\, h_{\mathrm{SD}} \, c(t)+ \sum_{m=1}^M \sqrt{\textsf{L}_{\mathrm{T}_m\mathrm{D}}}\,h_{\mathrm{T}_m\mathrm{D}} \, u_m(t) + n(t)\nonumber\\[3pt]
&= \sqrt{\textsf{L}_{\mathrm{SD}}}\,h_{\mathrm{SD}} \, c(t) + n(t)\nonumber \\
&\hspace{24pt}+ \sum_{m=1}^M \sqrt{\eta\,\textsf{L}_{\text{ST}_m} \textsf{L}_{\mathrm{T}_m\mathrm{D}} }\, h_{\text{ST}_m} \, h_{\text{T}_m\mathrm{D}}  \, \left[A_s-\Gamma_m(t)\right] \,  c(t) \nonumber \\
&=  \sqrt{2P} \left [\sqrt{g_0} \,h_0 + \sum_{m=1}^{M} \sqrt{g_m} \, h_m \, \mathcal{Y}_m(t) \right ] \, m(t) + n(t),
\label{eq:system}
\end{align} 
 where $n(t)$ is the thermal noise at the receiver and 
 \begin{align}
 g_0 &= \textsf{L}_{\mathrm{SD}},\\
 g_m &= \eta \,\textsf{L}_{\text{ST}_m} \textsf{L}_{\mathrm{T}_m\mathrm{D}}\, \mathbb{E}\left [ \left |A_s-\Gamma_m(t) \right|^2 \right]  ,\\
 \mathcal{Y}_m(t) & = \frac{A_s-\Gamma_m(t)}{\sqrt{ \mathbb{E}\left [ \left |A_s-\Gamma_m(t) \right|^2 \right] }},\\
 y_m\left[\Gamma_m(t)\right] &\stackrel{\triangle}{=} \sqrt{g_m}  \,h_m \, \mathcal{Y}_m(t).
 \end{align}

Notice that $\mathbb{E}[|h_0|^2]=\mathbb{E}[|h_m|^2]=\mathbb{E}[|  \mathcal{Y}_m  |^2]=1$ since $\mathbb{E}[|h_m|^2] = \mathbb{E}[|h_{\text{ST}_m} |^2] \, \mathbb{E}[|h_{\text{T}_m\mathrm{D}}|^2] $, due to the followed assumptions. It is also noted that $ \mathbb{E}\left [ \left |A_s-\Gamma_m(t) \right|^2 \right] = (1/K) \sum_{k=1}^K \left |A_s-\Gamma_k \right|^2. $

The above model is valid when coupling among the tags is negligible, i.e., the tags are separated by distance at least equal to $\lambda/2$.  Additive thermal noise $n(t)$ is modelled  by a complex, circularly symmetric, additive Gaussian noise process with  $\mathbb{E}[|n(t)|^2] = N_0 B$, where $B$ stands for receiver's bandwidth.\footnote{\mbox{$N_0=k_bT_{\theta}$}, where $k_b$ and $T_{\theta}$ are the Boltzmann constant and receiver temperature, respectively.} 
\section{Optimal Gain}
\label{sec:gain}

RIS targets at SNR improvement, by controlling the environment, through proper selection of the reflection coefficient at each element. According to Eq.~\eqref{eq:system}, the following instantaneous power maximization problem is formulated:
\begin{align}
 &\max_{\{ \mathcal{Y}_m(t) \}} \left | \sqrt{g_0} \,h_0 + \sum_{m=1}^{M} \sqrt{g_m} \, h_m \, \mathcal{Y}_m(t) \right |^2   2P \\
 &\max_{\{ \mathcal{Y}_m(t) \}} \left | \underbrace{\sqrt{g_0} \,h_0}_{y_0}+ \sum_{m=1}^{M} \sqrt{g_m} \, h_m \, \mathcal{Y}_m(t) \right | \\
        &=  \max_{\{ \Gamma_m(t) \}} \left| y_0 + \sum_{m=1}^{M} y_m\left[\Gamma_{m(t)} \right]  \right|, \label{eq:problem}
 \end{align}
which cannot be solved with exhaustive search, since each RIS element (among $M$ elements) can select among $K$ loads, i.e., $\Gamma_m(t)  \in \{ \Gamma_1, \Gamma_2, \ldots ,\Gamma_K \}$,  and thus, there are $K^M$ possible load configurations. Even for $K=2$ and $M=100$, exhaustive search among $2^{100}$ load configurations is not an option. 

The problem above is similar to noncoherent sequence detection of orthogonally-modulated sequences, solved with polynomial complexity in \cite{AlFouKarBl:16}. The trick is to introduce an auxiliary scalar variable $\phi$ into the problem of Eq.~\eqref{eq:problem}:
\begin{equation}
        \begin{split}       
        &\max_{\{ \Gamma_m(t) \}} \max_{\phi \in [0,2 \pi) } \Re\left\{ e^{-j\phi}\left( y_0 + \sum_{m=1}^{M} y_m\left[\Gamma_{m(t)} \right] \right) \right\} = \\ 
        &\max_{\phi \in [0,2 \pi) } \max_{\{ \Gamma_m(t) \}} \Bigg(\Re\left\{e^{-j\phi} y_0\right\} + \sum_{m=1}^{M}\Re\left\{e^{-j\phi} y_m\left[\Gamma_{m(t)} \right]\right\}\Bigg) \label{eq:problem2}
        \end{split}
\end{equation}

\subsection{$K = 2$ Loads}
For a given point $\phi \in [0,2\pi)$, the innermost maximization in Eq.~\eqref{eq:problem2} is separable for each $\Gamma_{m}(t)$ and hence, splits into independent maximizations for any $m=1,2,\dots,M$:
\begin{align}
\hat{\Gamma}_{m}(t)&=\underset{\Gamma_m(t) \in\{\Gamma_1,\Gamma_2\}}{\arg \max } \,\Re\left\{\mathrm{e}^{-j \phi} y_{m}\left[\Gamma_{m}(t)\right]\right\} \cr
& \Leftrightarrow  \Re\left\{\mathrm{e}^{-j \phi} y_{m}\left[\Gamma_1\right]\right\}  \underset{\hat{\Gamma}_{m}(t) = \Gamma_2 }{\overset{\hat{\Gamma}_{m}(t) = \Gamma_1}{\gtrless}}  \Re\left\{\mathrm{e}^{-j \phi} y_{m}\left[\Gamma_2\right]\right\} \cr
& \Leftrightarrow  \Re\left\{\mathrm{e}^{-j \phi} \left(y_{m}\left[\Gamma_1\right]-y_{m}\left[\Gamma_2\right]\right)\right\}  \underset{\hat{\Gamma}_{m}(t) = \Gamma_2 }{\overset{\hat{\Gamma}_{m}(t) = \Gamma_1}{\gtrless}}  0 \cr
& \Leftrightarrow \cos\left(\phi -\phase{y_{m}\left[\Gamma_1\right]-y_{m}\left[\Gamma_2\right]}\right) \underset{\hat{\Gamma}_{m}(t) = \Gamma_2 }{\overset{\hat{\Gamma}_{m}(t) = \Gamma_1}{\gtrless}}  0
\label{eq:cos_phi}
\end{align}

Given the relation in Eq.~\eqref{eq:problem2}, the optimal load sequence $\hat{\mathbf{\Gamma}}^{\text{opt}}$ can be found by varying $\phi$ from $0$ to $2\pi$. 
 It is further noticed that, as $\phi$ scans $\left[0,2\pi \right)$, the decision $\hat{\Gamma}_{m}(t)$ changes, according to Eq.~\eqref{eq:cos_phi}, only when:
\begin{equation}
    \begin{split}
    &\cos\left(\phi - \phase{y_{m}\left[\Gamma_1\right]-y_{m}\left[\Gamma_2\right]}\right) = 0 \cr
    &\Leftrightarrow \phi =
    \underbrace{\pm\frac{\pi}{2} + \phase{y_{m}\left[\Gamma_1\right]-y_{m}\left[\Gamma_2\right]} \quad \left(\operatorname{mod} 2\pi\right)}_{\phi^{(1)}_m,\phi^{(2)}_m}.
    \end{split}
\end{equation}

Therefore, the sequence $\hat{\mathbf{\Gamma}} = \left [\hat{\Gamma}_1(t), \hat{\Gamma}_2(t),\dots,\hat{\Gamma}_M(t)\right]^T$ changes only at $\left(\phi^{(1)}_1,\phi^{(2)}_1,\phi^{(1)}_2,\phi^{(2)}_2,\cdots,\phi^{(1)}_M,\phi^{(2)}_M\right)$. For the remaining part of this section, we assume that the above $2M$ points are distinct and nonzero, i.e.,  $\phi^{(j)}_m \neq \phi^{(k)}_l$ and $\phi^{(j)}_m \neq 0$, for any $j,k, \in \{1,2\}$ and $m,l \in \{1,2,\dots,M\}$ with $m\neq l$. There is a case where the above assumption does not hold, examined in \cite{AlFouKarBl:16}. If the above points are sorted in ascending order, i.e.,
\begin{align}
& \left(\theta_1,\theta_2,\cdots,\theta_{2M}\right) = \nonumber \\
 &=\textsf{sort} \left(\phi^{(1)}_1,\phi^{(2)}_1,\phi^{(1)}_2,\phi^{(2)}_2,\cdots,\phi^{(1)}_M,\phi^{(2)}_M\right), 
 \label{eq:sort}
\end{align}
then the decision $\hat{\mathbf{\Gamma}}$ will remain constant in each one of the $2M+1$ intervals $(\theta_i, \theta_{i+1}), i \in \{0, 1, \ldots, 2M  \}$, with $\theta_0=0$ and $\theta_{2M+1}=2\pi$. The goal is the identification of the $2M+1$ sequences that correspond to these intervals,\footnote{It can be shown that the sequence at $[0,\theta_1)$ is the same with the sequence at $[\theta_{2M}, 2\pi)$ and thus, $2M$ intervals/sequences should be identified.} one of which gives the optimal $\hat{\mathbf{\Gamma}}^{\text{opt}}$, i.e., the one that offers the maximum power; thus, the quality of each sequence is calculated with the norm metric of Eq.~\eqref{eq:problem}, which explicitly includes the direct 
channel $h_0$. Based on the above, the sorting operation in Eq.~\eqref{eq:sort} is dominant in terms of computational cost, which is $\mathcal{O}(M\log{M})$ and not $2^M$.
%
  
\subsection{$K>2$ Loads}
\label{sec:gain_K}
The method described above can be generalized to $K>2$ loads, i.e.,  $\Gamma_m(t)$ belongs in
$\{ \Gamma_1, \Gamma_2, \ldots ,\Gamma_K \}$. The solution is given by selecting the largest value of $\Re\left\{\mathrm{e}^{-j \phi} y_m[\Gamma_k]\right\}$ among all $k \in \{1,2, \ldots K\}$, which results in testing \mbox{$2M \times (K-1)$} changes of $\phi$ and as a result, same number of sequence changes and not $2M \times {K \choose 2}$, as one would expect; the rest of the steps are exactly the same as in $K=2$. Formal proof and details 
can be found in \cite{AlFouKarBl:16}, omitted due to space constraints. Notice that the norm metric for the quality of each sequence must include 
$h_0$. The complexity of the algorithm is again $\mathcal{O}(M\log{M})$ for $M>K$ and not $K^M$.
 
\begin{figure}[!t]
    \centering
	\includegraphics[width=0.8\columnwidth]{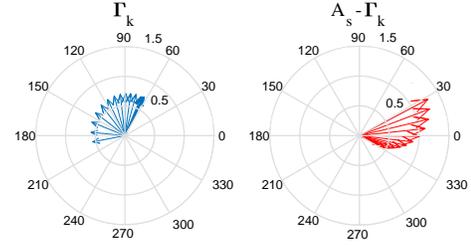}
    \caption{The $21$ reflection coefficients used in the simulations.}
    \label{fig:loads}
\end{figure}

Fig. $\ref{fig:loads}$-left depicts $K=21$ (complex) reflection coefficients, corresponding to the $21$ loads, offered through a varactor at each tag, as experimentally tested in \cite{VouBleSah:20}. It can be shown that these reflection coefficients span more than $120^{\circ}$. Fig. $\ref{fig:loads}$-right shows $A_s - \Gamma_k, k \in \{1,2, \dots K \}$, which incorporates the contribution of the structural mode of each RIS element antenna, typically overlooked in the literature; it can be seen that the span of  $A_s - \Gamma_k$ in  degrees is much smaller, in the order of $60^{\circ}$.

\section{Wireless Batteryless Implementation}

The RFID-based RIS is controlled by the RFID reader at carrier frequency $f_2$, through commands such as Select, Query and ACK, explained below. In the commercial RFID standard (EPC Gen2), a  framed  slotted  Aloha (FSA) protocol is used, so that tags backscatter one at a time, without collision. For RIS purposes, the opposite is needed, i.e., tags must be forced 
to backscatter in carefully selected groups. 
\begin{figure}[!ht]
        \captionsetup[subfigure]{justification=centering}
		\subfloat[No RN16 collision.
		]{\includegraphics[width=0.5\columnwidth]{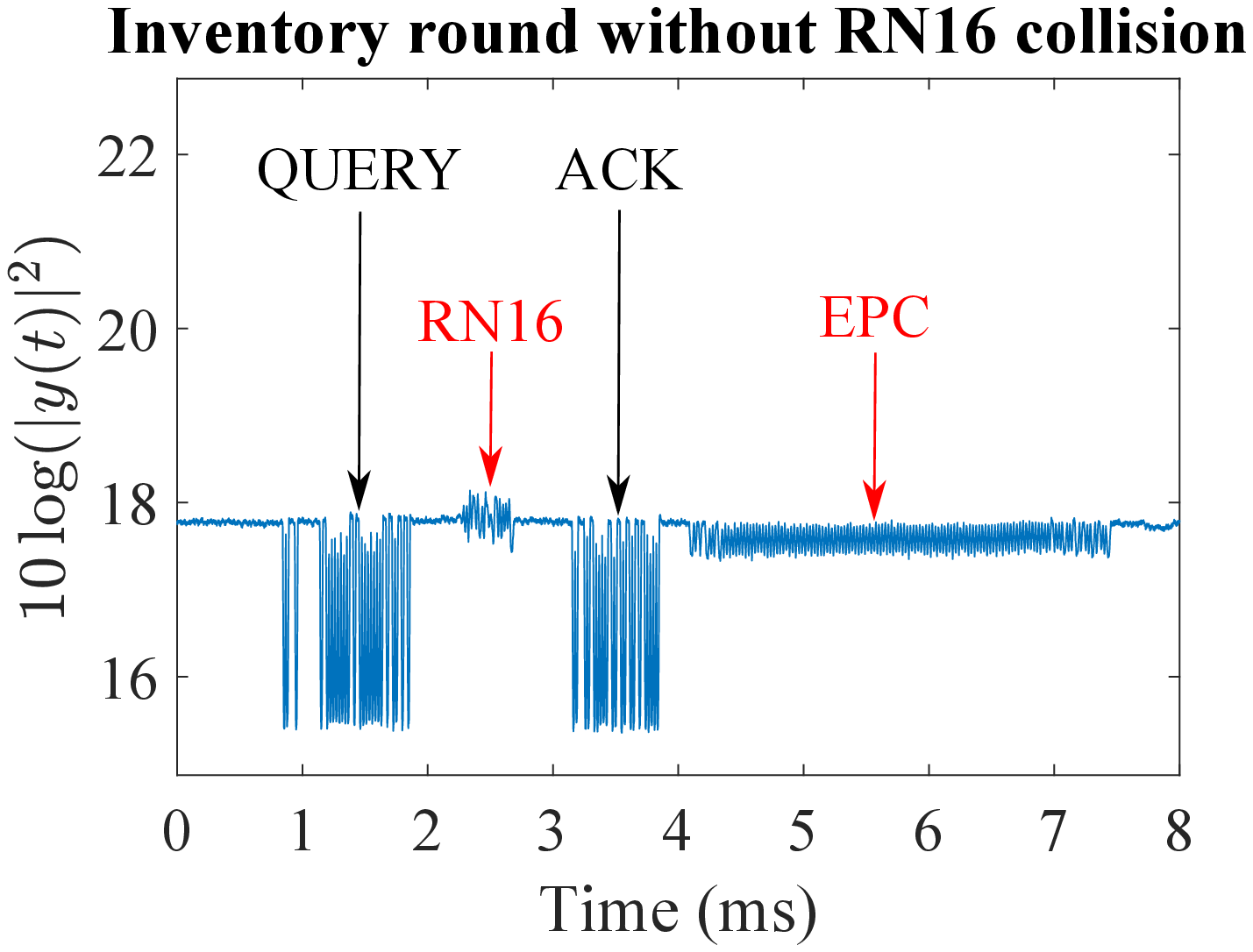}}%
		\subfloat[RN16 collision.
		]{\includegraphics[width=0.5\columnwidth]{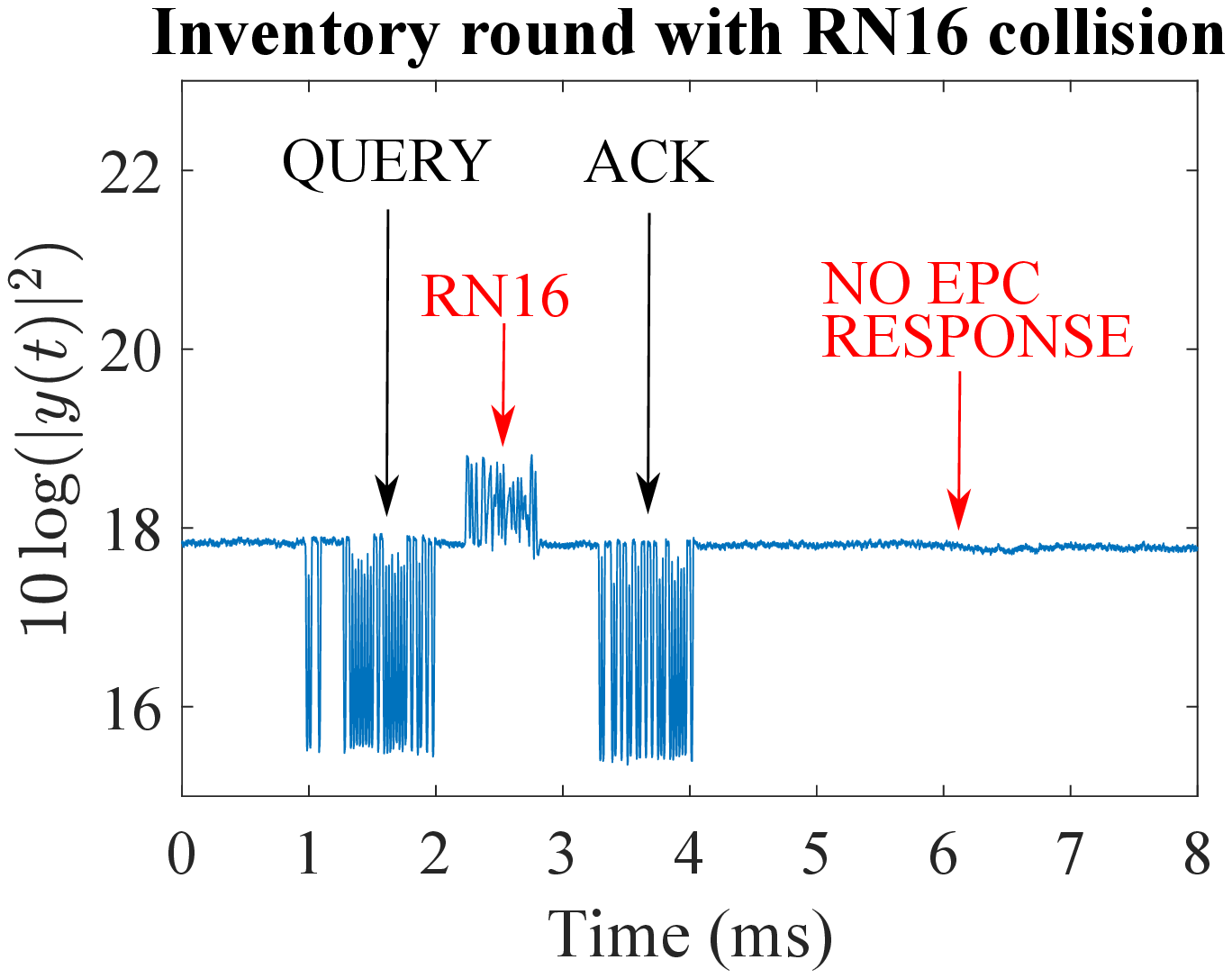}}%
		\caption{Conventional Gen2 RFID operation: each tag responds to a Query 
command with a random 
        RN16 message; if it is correctly acknowledged it will respond with its EPC.}
		\label{fig:epc}
\end{figure}

In Gen2, the reader initiates the start of a frame with a Query command, which contains the number of slots. Ideally, the number of slots should be equal to the number of tags to be inventoried. If the reader advertises 
number of slots equal to $1$, then all tags in the vicinity of the reader 
will respond. That was the approach followed in this work. Then, each tag 
responds with a random $16$-bit number, namely the RN16 message, which is 
in principle different among the competing tags. If the reader correctly decodes that message then it will send the ACK command, containing the RN16 sent by the tag. Error detection is conducted at the reader using the line coding the tags incorporate (FM0 or Miller).\footnote{The reader directs the tags about  the line code they are going to use.} If the tag is correctly acknowledged it will reply with its ID (EPC); the latter is typically $96-$bits payload plus CRC bits. Fig.~\ref{fig:epc}-left depicts an inventory round with an acknowledged tag, while in Fig. \ref{fig:epc}-right multiple tags have sent their RN16 resulting to a collision; this is the case exploited in this work.


\begin{figure}[!hb]
    \centering
	{\includegraphics[width=0.7\columnwidth]{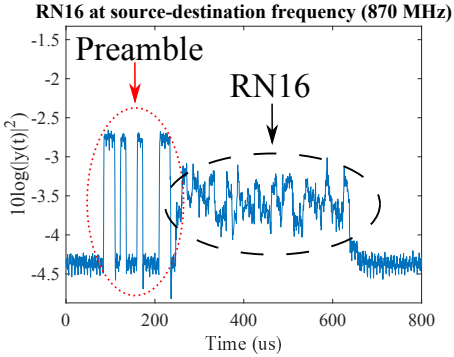}}
    \caption{Superposition of Preamble$+$RN16 from multiple tags.}
    \label{fig:es6}
\end{figure}

The RN16 is preceded by a $6$-bit Preamble sequence, which does not follow the line coding rules and is the same for all tags. As a result, using the Preamble one could measure the effect of a specific RIS \emph{configuration} (where configuration means the set of tags that change their reflection coefficients, among the total number of RIS tags). Since the random $16$-bit sequence of the RN16 is different for each tag, the signal (or 
the effective channel) during RN16 could be higher or lower than the Preamble because only a subset of the configuration's tags are terminated at the same load at a given point of time. This is shown in Fig. \ref{fig:es6}.

\begin{figure}[!t]
    \centering
	{\includegraphics[width=0.7\columnwidth]{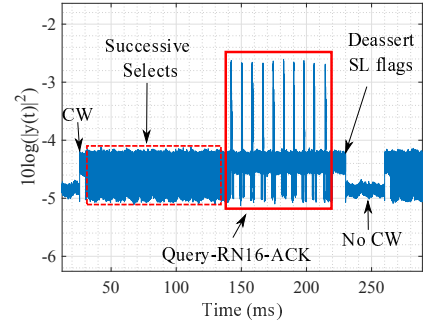}}
    \caption{Reader-Tag communication during a configuration.}
    \label{fig:suc_sel}
\end{figure}

The Select command is asserting or deasserting the SL flag of the tags. When a tag has its SL flag asserted, it responds with an RN16 after a Query command. The Select command contains a mask with a specified length and 
starting point. Each tag compares this mask with its corresponding EPC bits. The Action parameter defines how the SL flag will change if the mask matches the EPC of the tag. To assert a specific tag's SL flag without affecting the others', the Action parameter is set to $\{0,0,1\}$. To deassert all SL flags with one Select, the Action parameter is set to $\{0,0,0\}$ and the mask is set to avoid corresponding to any tag.

\begin{figure}[!hb]
    \centering
	{\includegraphics[width=0.7\columnwidth]{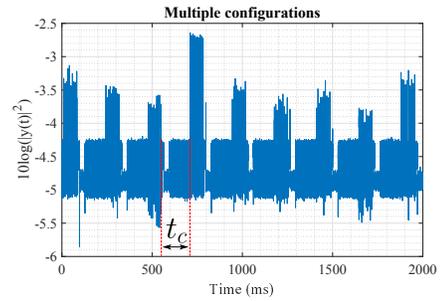}}
    \caption{Power at Destination during different configurations.}
    \label{fig:es5}
\end{figure}

For RIS purposes, a \emph{specific} configuration is set by having the reader issuing successive Select commands and asserting the SL flag of the specified tags, sequentially. This is shown in Fig.~\ref{fig:suc_sel}. First, a continuous wave (CW) energizes the tags and then the Select commands are issued. Afterwards, we can observe $10$ inventory rounds (Query-RN16-ACK), where the peaks are the superposition of the RN16 from the selected tags. Finally, to advance to a new configuration, another Select 
is sent but with Action parameter equal to $\{0,0,0\}$ and a non-matching 
mask to any tag.

In Fig.~\ref{fig:es5} we observe a sequence of $9$ different  configurations. Each configuration spans $10$ inventory (Query-RN16-ACK) rounds. Different configurations offer different effective channel and hence, different maximum signal power. Those peaks occur at the Preamble+RN16 tag response, as explained in Fig.~\ref{fig:suc_sel}. 
It is noted that the measurements were conducted in a static indoor environment, with channel coherence time spanning several hundreds of msec (and thus, different peaks are due to different tag configurations and not changes in the channel). The time between the last RN16 from one configuration and the first RN16 from the next is specified as $t_c$. More specifically, by varying the backscatter link frequency (BLF) parameter, $t_c$ changes accordingly; for example, if $\mu=50$ active tags are selected (among a population of $M\geq \mu$), then $t_c \in \left[43 \text{ ms},116\text{ ms} \right]$. In Fig.~\ref{fig:es5}, $t_c=116$ ms. Future RIS-friendly modifications of Gen2 could further reduce this parameter. 
\begin{figure}[!t]
    \centering
	{\includegraphics[width=0.8\columnwidth]{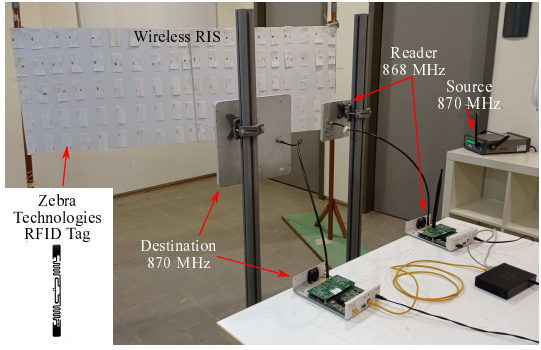}}
    \caption{Experimental setup.}
    \label{fig:es7}
\end{figure}

The experimental setup is depicted in Fig.~\ref{fig:es7}, with $\text{M}=100$ commodity Zebra Z-Perform $1500\text{T}$ Gen2 RFID tags , separated by $d_x=10$ cm and $d_y=5$ cm. The source-destination link operates at $f_1=870$ MHz, while the software-defined radio (SDR) reader 
operates at $f_2=868$ MHz, with zero interference among the two channels. Modifications in the Gen2  reader software stack from \cite{KaMaBl:15} 
were conducted to enable RIS functionality. Two commodity USRP N200 software defined radios (SDR), each equipped  with a RFX900 daughterboard, were utilized, both connected to a commodity laptop through a switch. The antennas used at the reader and Source were circularly-polarized. It is emphasized that Figs.~\ref{fig:epc}, \ref{fig:es6}, \ref{fig:suc_sel}, \ref{fig:es5} are offered from the SDR receiver at the destination. 

\section{Numerical Results}
\subsection{Simulations}
\begin{figure}[!t]
    \centering
	{\includegraphics[width=0.8\columnwidth]{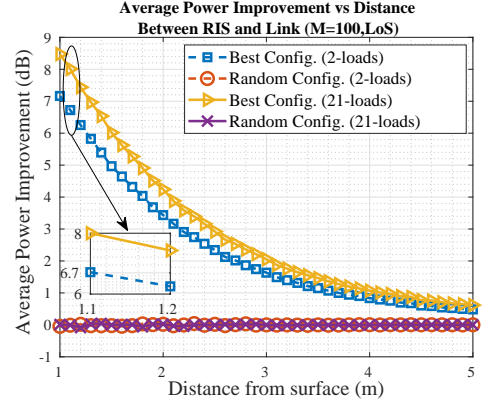}}
    \caption{Average Power Gain; tags' spacing $d_x=d_y=\frac{\lambda}{2}$.}
    \label{fig:gain_vs_distance_LOS}
\end{figure}

\begin{figure}[!t] 
        \captionsetup[subfigure]{justification=centering}
		\subfloat[
		]
		{\includegraphics[width=4.5cm]{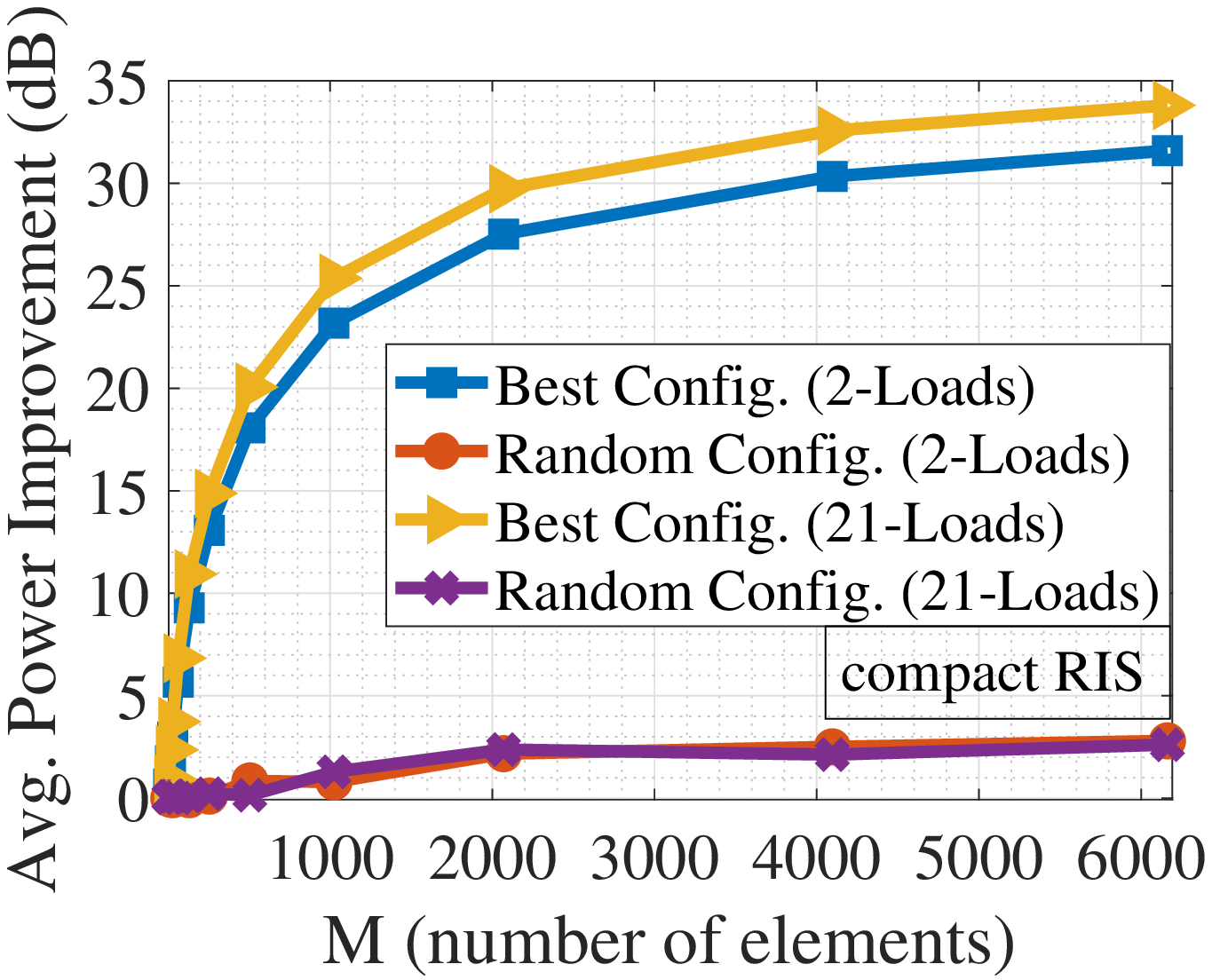}}%
		\subfloat[
		]{\includegraphics[width=4.5cm]{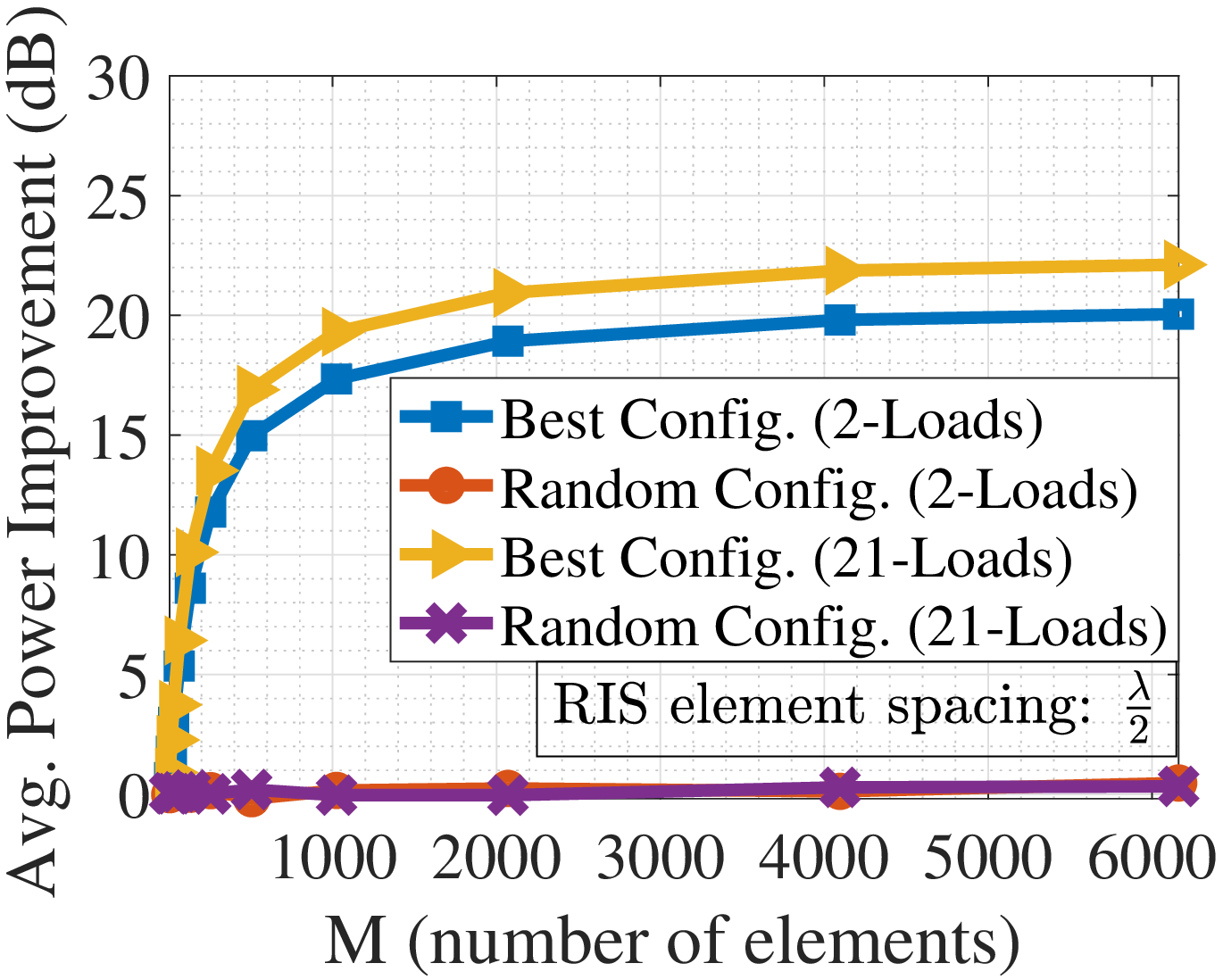}}%
		\caption{Average Power Improvement vs M (direct link present).}
		\label{fig:gain_vs_M_LOS}
\end{figure}

\begin{figure}[!tb]
        \captionsetup[subfigure]{justification=centering}
		\subfloat{\includegraphics[width=4.5cm]{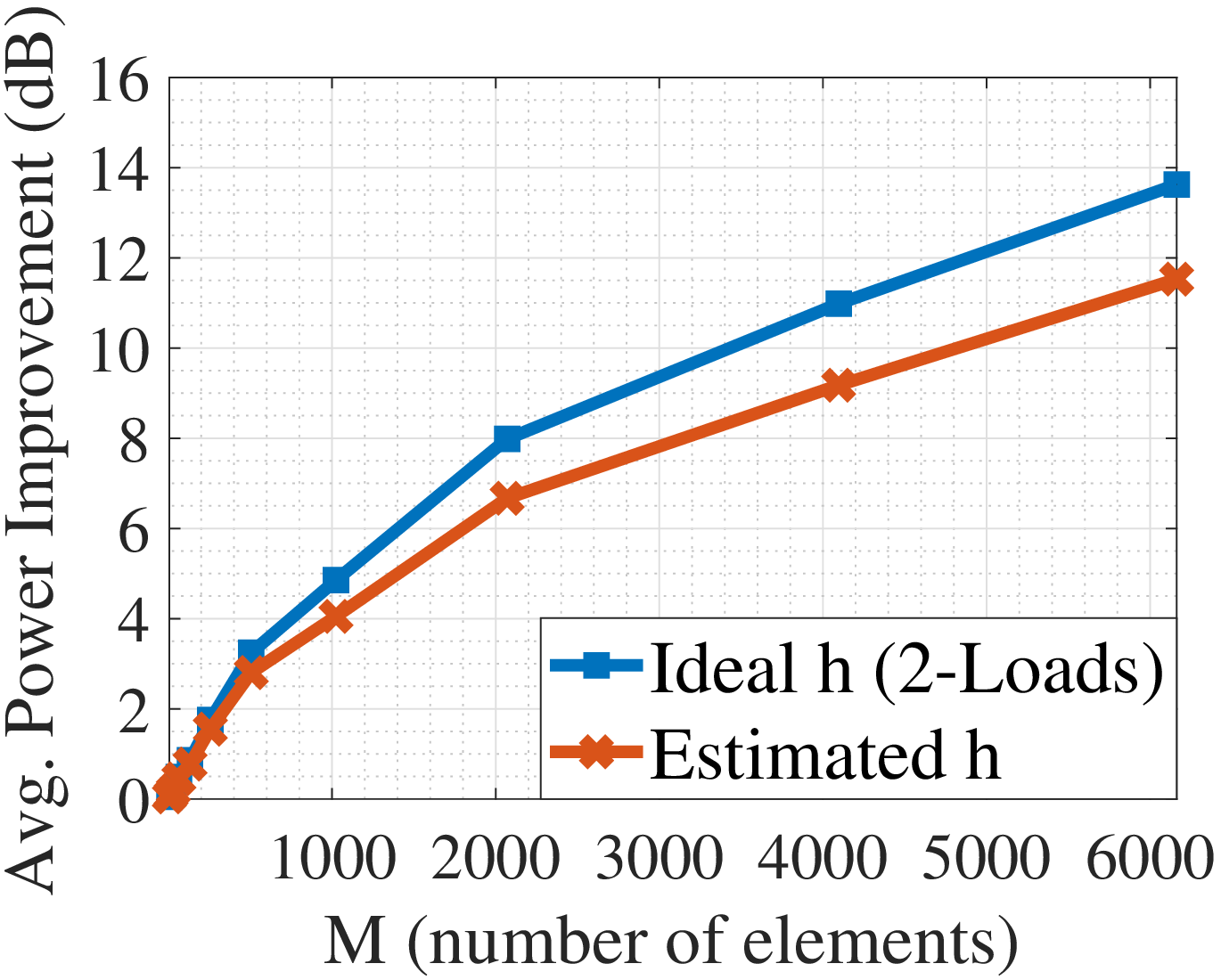}}
		\subfloat{\includegraphics[width=4.5cm]{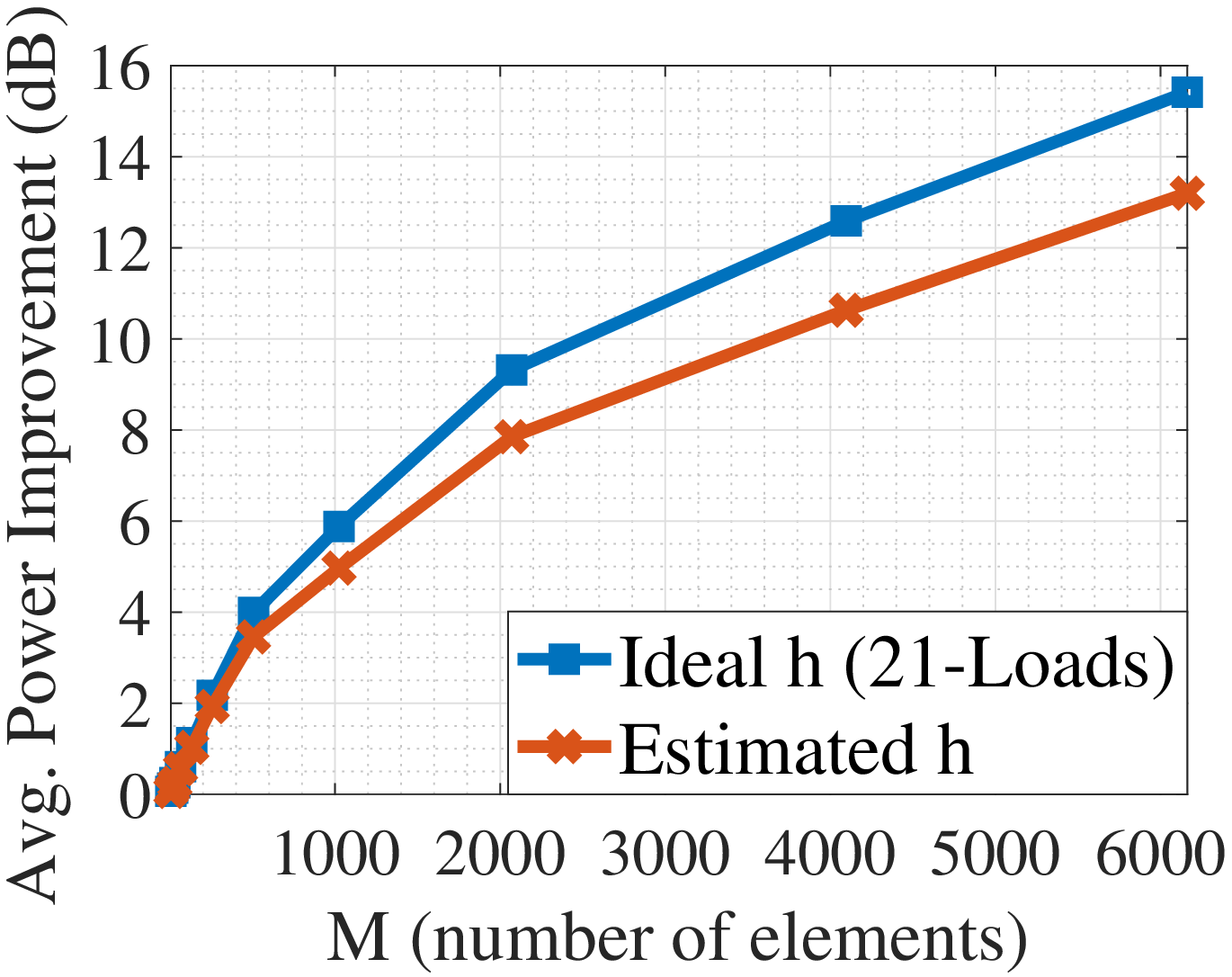}}
		\caption{Impact of Channel Estimation Error.}
		\label{fig:CSIest}
\end{figure}

Figs.~\ref{fig:gain_vs_distance_LOS} and \ref{fig:gain_vs_M_LOS} are generated with typical indoor static 
conditions in mind: $\kappa_{\mathrm{S}\mathrm{T}_m}=\kappa_{\mathrm{T}_m\mathrm{D}}=\kappa_{\mathrm{SD}}=8$, $\eta = 10\%$, $v_\mathrm{X}=3$, $d_0^{\mathrm{X}}=d_{\mathrm{SD}}=3$ m, $f_2=870$ MHz, $P=5$ dBm ($3.16$ mW), and $10$ dB relative end-2-end antenna gain for the 
backscattered links compared to direct link, assuming that the source and 
destination antennas point towards the RIS (in order to assist its operation). The same parameters hold for Fig.~\ref{fig:CSIest}, unless otherwise noted.

Fig.~\ref{fig:gain_vs_distance_LOS} offers the average power improvement due to RIS and direct link operation, compared to power of the direct link communication. Several instantiation of the channels are generated, optimal configuration for $K=2$ or $K=21$ is found, based on the analysis 
of Section~\ref{sec:gain} and average values are reported. The SD link is parallel to RIS and distance from surface ($d_{\mathrm{RIS-SD}}$) is varied. It is observed that as the distance of the SD link from RIS increases, performance gains to optimal operation of RIS decrease. That is due to the fact that backscattered links (through RIS) become weaker, as the SD link is farther away from the surface. 
This finding suggests that amplification at each RIS element (i.e., $| \Gamma_k | >1$) is needed. Finally, it is observed that using $K=21$ loads for each element, despite the small angle span of the induced backscattered signals (explained in Sec.~\ref{sec:gain_K}), offers power gain in the order of  $1.3$ additional dB, compared to $K=2$.

Fig.~\ref{fig:gain_vs_M_LOS} offers average power improvement as a function of $M$, for two setups: Fig.~\ref{fig:gain_vs_M_LOS}-left sets $d_x=0.1$ m and $d_y=0.05$ m (setup (a)), while  Fig.~\ref{fig:gain_vs_M_LOS}-right sets spacing between adjacent tags equal to $\lambda/2$ (setup (b)). For both setups, $d_{\mathrm{SD}}=3$ m, $d_{\mathrm{RIS-SD}}=1$ m, 
while setup (a) ignores possible coupling between adjacent tags. It can be noticed that performance in (b) reaches a plateau with a faster rate than in (a), as a consequence of the larger element 
spacing. This finding suggests that even with perfect channel estimation, the weak nature of backscattered links limits the performance gains, even for large number of tags/RIS elements.

Finally, Fig.~\ref{fig:CSIest} quantifies the impact of channel estimation error $\tilde{h}_m$ on the estimation of $h_m$, with $\hat{h}_m$ being the channel estimate, i.e., $h_m = \hat{h}_m + \tilde{h}_m$, assuming minimum channel estimation error (MMSE) estimators, with $\mathbb{E}[| \tilde{h}_m |^2] = \textsf{MMSE}_m$ and expressions for $\textsf{MMSE}_m, m \in \{0, 1,2, \ldots M\}$ from the multi-antenna literature \cite{HaHo:03}, \cite{LoAnHeRoAnJe:13}. Rayleigh fading is assumed ($\kappa_{\mathrm{S}\mathrm{T}_m}=\kappa_{\mathrm{T}_m\mathrm{D}}=\kappa_{\mathrm{SD}}=0$), with $v_\mathrm{X}=3$, $d_{\mathrm{RIS-SD}}=8$ m, $d_0^{\mathrm{X}}=3$ m, $d_{\mathrm{SD}}=15$ m and $B=48$ MHz. The optimal configuration is found with estimated channels $\{ \hat{h}_m\}$, while the effective channel and gains are found based on the 
true channels and the offered configurations. Channel coherence time in number of symbols is set to $L_c = 24 \times 10^5$, corresponding to channel coherence time of $100$ ms and SD link at $48$ Mbps with QPSK modulation. If $\alpha$ is the percentage of $L_c$ devoted to pilot transmission for channel estimation then  $\alpha \, L_c > M$, since the number of pilot symbols cannot be less than the number of 
unknown channels; in this plot $\alpha$ varies from $1\%$ to $11\%$ as $M$ increases. It is found that channel estimation error for large number of elements $M$ has important, non-negligible impact.

\subsection{Experiments}
\begin{figure}[!t]
        \captionsetup[subfigure]{justification=centering}
		{\includegraphics[width=0.5\columnwidth]{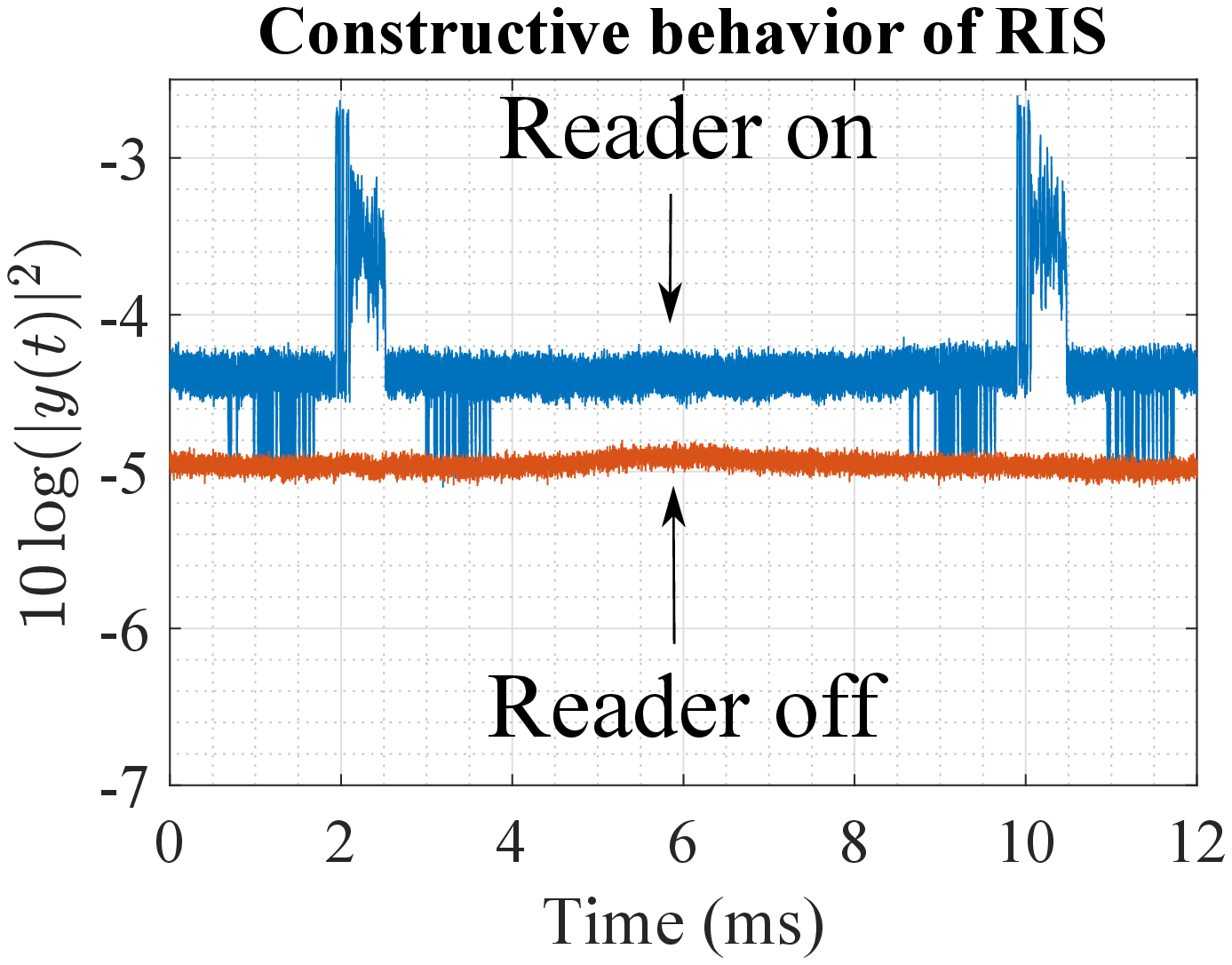}}%
		{\includegraphics[width=0.5\columnwidth]{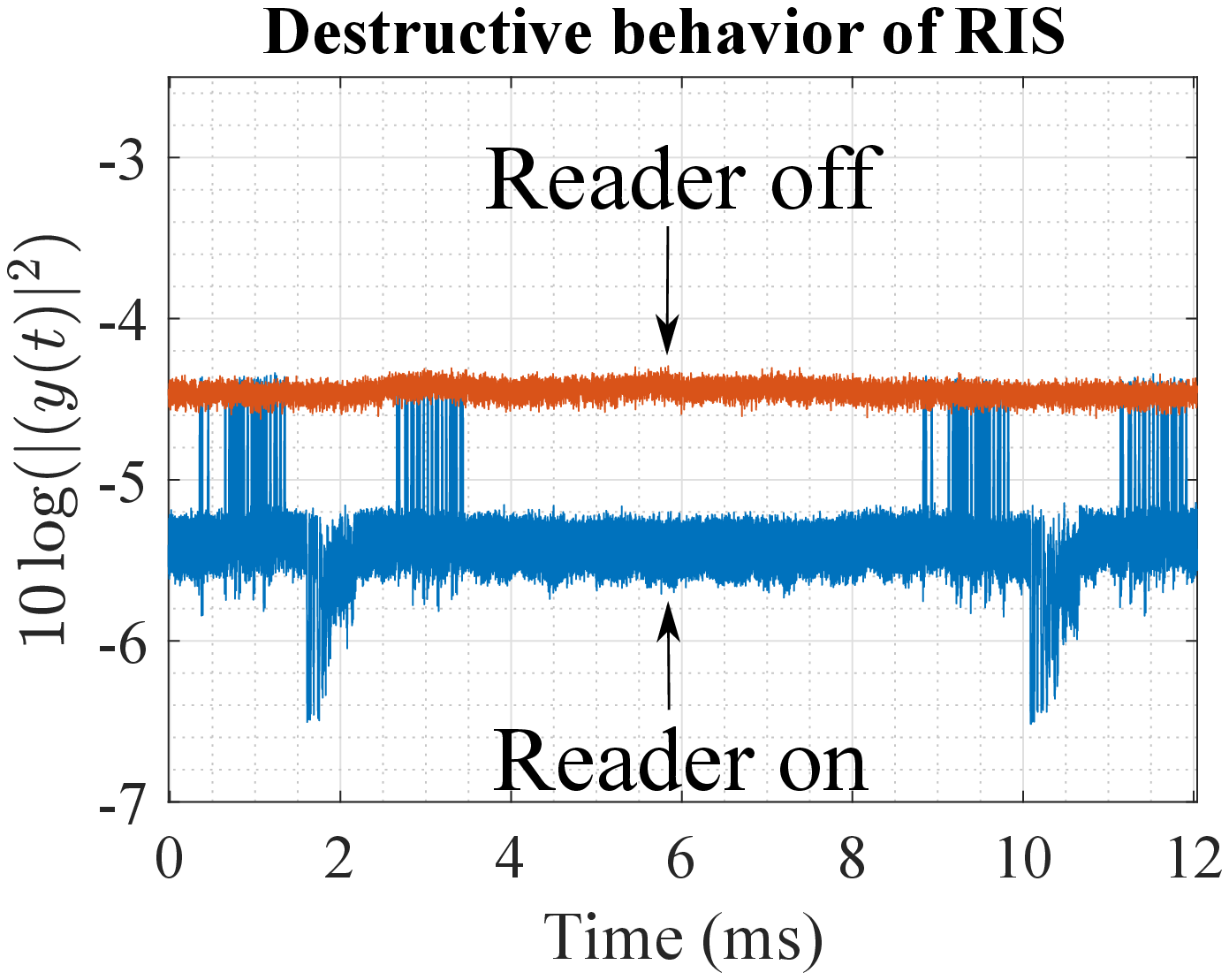}}%
        \caption{Constructive vs destructive RIS impact at Destination.}
        \label{fig:con_des}
\end{figure}

Fig.~\ref{fig:con_des} offers the received power at the destination, operating @ $870$ MHz, for $2$ different setups of Fig.~\ref{fig:es7}. With the reader on (@ $868$ MHz), both the RIS and the SD link contribute to the received power (@ $870$ MHz). With the reader switched off, the received power at the destination is measured and depicted. Clearly, one setup offers constructive and another destructive RIS operation. The source-RIS distance was denoted as $d_{\mathrm{RIS-S}}$ and the destination-RIS as $d_{\mathrm{RIS-D}}$, measured from the RIS center; $\phi_\mathrm{SD}$ is the angle between the direction of the destination antenna and the SD link. For the constructive setup,  \mbox{$d_{\mathrm{RIS-S}}=2.5$ m}, $d_{\mathrm{RIS-D}}=1.1$ m, $\phi_\mathrm{SD}=100^{\circ}$ and $d_{\mathrm{SD}}=2.1$ m. For the destructive setup, $d_{\mathrm{RIS-S}}=2.1$ m, $d_{\mathrm{RIS-D}}=1.1$ m, \mbox{$\phi_\mathrm{SD}=80^{\circ}$}, $d_{\mathrm{SD}}=2$ m.

\begin{figure}[!t]
    \centering
	{\includegraphics[width=0.75\columnwidth]{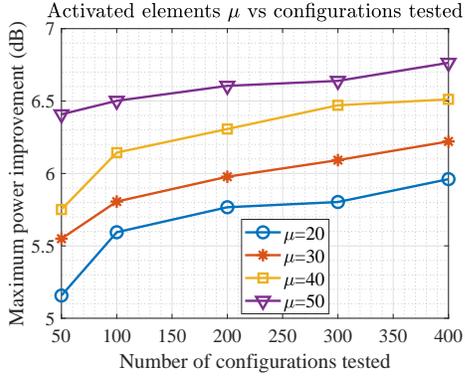}}
    \caption{Maximum power improvement vs Number of configurations tested for several $\mu$. }
    \label{fig:es9}
\end{figure}

Fig. \ref{fig:es9} shows the maximum power improvement achieved as a function of number of configurations tested, for variable number of activated tags (denoted by  $\mu$). The setup corresponds to \mbox{$d_{\mathrm{RIS-S}}=2.2$ m}, $d_{\mathrm{RIS-D}}=0.9$ m, $\phi_\mathrm{SD}=145^{\circ}$ and $d_{\mathrm{SD}}=1.4$ m. For a given configuration and $\mu$, the value reported is over three repetitions. Increasing the number of configurations tested results to an increased maximum power value. Also, when $\mu$ is increased, the maximum gain is increased for a given number of configurations. The increase of the power improvement in Fig. \ref{fig:es9} at $6.5$ dB versus the $2.5$ dB of Fig. \ref{fig:con_des}-left is due to the change in the antenna gain for different $\phi_\mathrm{SD}$, since the destination antenna is directive, affecting the SD link's contribution.  

\balance
\bibliographystyle{IEEEtran}
\bibliography{IEEEabrv,BLETSAS_GROUP_bib_v63.bib}
\end{document}